\documentclass[useAMS,fleqn,usenatbib]{mnras}  

\usepackage{newtxtext,newtxmath}
	
\usepackage[T1]{fontenc}
\usepackage{ae,aecompl}

\usepackage{graphicx}	
\usepackage{amsmath}	
\usepackage{amssymb}	
\input psfig.sty       
\RequirePackage{xcolor}

\newcommand{\lt}{\symbol{"3C}}
\newcommand{\gt}{\symbol{"3E}}
\def \grss {GRS 1915+105~}
\def \grs {GRS 1915+105}

\def \rxx {{\it RXTE}~}

\defcitealias{Massaro2020}{Paper~II}

\title[A non-linear model for the variability classes of GRS~1915+105 - I]
{A non-linear mathematical model for the X-ray variability classes of the 
microquasar GRS~1915+105 - I: quiescent, spiking states and QPOs} 

\author[E. Massaro, F. Capitanio, M. Feroci, T. Mineo, A. Ardito, P. Ricciardi]
  {E. Massaro$^{1}$,
  F. Capitanio$^{1}$, 
  M. Feroci$^{1}$,
  T. Mineo$^{2}$\thanks{E-mail address: \texttt{teresa.mineo@inaf.it}},
  A. Ardito$^{3}$, 
  P. Ricciardi$^{3}$  \\
$^1$ INAF, IAPS, via del Fosso del Cavaliere 100, I-00113 Roma, Italy \\
$^2$ INAF, IASF Palermo, via U. La Malfa 153, I-90146 Palermo, Italy \\
$^3$ Sapienza Universit\`a di Roma, Piazzale Aldo Moro, 5, I-00185 Roma, Italy \\
}
      
\date{}    

\pubyear{2020}

\begin{document}

\label{firstpage}
\pagerange{\pageref{firstpage}--\pageref{lastpage}}
\maketitle

\begin{abstract}
The microquasar \grss is known to exhibit a very variable X-ray emission 
on different time scales and patterns.
We propose a system of two ordinary differential equations, adapted from the 
Hindmarsh-Rose model, with two dynamical variables $x(t)$, $y(t)$ and an input 
constant parameter $J_0$, to which we added a random white noise, whose solutions 
for the $x(t)$ variable reproduce consistently the X-ray light curves of several 
variability classes as well as the development of low frequency Quasi-Periodic 
Oscillations (QPO). 
We show that changing only the value of $J_0$ the system moves from stable to 
unstable solutions and the resulting light curves reproduce those of the quiescent
classes like $\phi$ and $\chi$, the $\delta$ class and the spiking $\rho$ class. 
Moreover, we found that increasing the values of $J_0$ the system induces high 
frequency oscillations that evolve to QPO when it moves into another stable region.
This system of differential equations gives then a unified view of the variability 
of \grss in term of transitions between stable and unstable states driven by a single 
input function $J_0$.
We also present the results of a stability analysis of the equilibrium points
and some considerations on the existence of periodic solutions.
\end{abstract}



\section{Introduction}
 
\grs, the first microquasar, discovered by \citet{CastroTirado1992}
is known to exhibit a large number of variability patterns on different 
time scales.
On long time scales, months to years, changes of the X-ray flux up to
about one order of magnitude are reported \citep{Huppenkothen2017} and
on time scales of thousands of seconds, light curves present either 
quiescent states, with red noise power spectra, or series of fast bursts. 
A first classification of light curves was performed by \citet{Belloni2000}
who defined 12 variability classes on the basis of the spectral and timing 
properties of a large collection of multiepoch observations.
New classes were discovered in the subsequent years \citep{KleinWolt2002,
Hannikainen2003, Hannikainen2005} indicating that \grss is potentially able 
to develop physical conditions from which new classes of light curves can 
be originated.
A useful compilation of light curves for about all classes can be found
in \citet{Polyakov2012}, and a more complete 
description of the rich phenomenology of this unique source is given in 
the review paper by \citet{Fender2004}.

The complex hydrodynamical, thermal and magnetic phenomena occurring in accretion
discs around black holes involve non-linear processes whose evolution can be 
described by a system of differential equations.
These can be solved by numerical calculations involving several quantities not 
directly observable, as the gas density or viscous stresses.
The stability of disc structures is also a very interesting subject 
of investigations since many years and theoretical analysis suggested that 
thermal and viscous instabilities can develop and establish a limit cycle 
behaviour.
In particular the possibility of observing a bursting behaviour originated
by thermal relaxtion oscillations between standard and slim disc states
was first suggested by \citet{Honma1991} and after by 
\citet{Chen1993, Szuszkiewicz1998}, before the discovery
of this behaviour in \grs.
Complexity of hydrodynamic and thermodynamic equations, however, does not 
allow a rather simple picture of the roles played by the involved physical 
quantities and the interpretation of data is not simple.
Moreover, the limit cycle is often described in terms of disc quantities, such 
as the surface density or the mass accretion rate, which are not directly 
observable.
As stated by \citet{Fender2004} in their review paper ``it will not be possible
to interpret in detail every aspect of the complex light curves ..., yet the
structure is not random and contains information about the accretion disc of
\grs.''

It has been noticed that the nearly regular burst sequences of \grss have some
similarities with signals from living systems and, in fact, it is frequently 
referred in the literature as the `hearthbeat' state \citep{Neilsen2011}.
There are many mathematical tools developed for describing similar behaviours
and particularly those for the bursting of neurons.
This analogy suggested us to search if some of these models could be applied
in the case of \grss and in a couple of previous papers \citep{Massaro2014,Ardito2017} 
we studied the solutions of a non-autonomous and non-linear system of two ordinary 
differential equations (ODE), whose solutions are able to reproduce the light curves 
of some variability classes.
More precisely, in those two papers we considered the ODE system introduced by 
\citet{Fitzhugh1961} and \citet{Nagumo1962} to model the quiescent and spiking 
behaviours of the classes $\phi$, $\chi$ and $\rho$.
In the present paper, we propose another system of two ODEs, based on that developed by 
\citet{Hindmarsh1984}, which is able to reproduce light curves of several 
variability classes that must be considered as a useful analytical approximation 
for describing the evolution of instabilities and the transitions between different 
equilibrium states.
Our goal is to clear up the most relevant features of the observed behaviours with 
the low complexity of the mathematical model: it will be possible
to show that some dynamical regimes, like quiescent or bursting
activity, are indeed obtained using a low-dimensional system of ODEs.
Moreover, we will show that non-linear processes provide a unified view of other 
interesting phenomena as the development of low frequency Quasi-Periodic Oscillations 
(QPO) in accretion discs.

Non-linear oscillators were already considered in a seminal paper by 
\citet{Moore1966} in a stellar physics context for describing the convective 
energy transfer.
Later, \citet{Usher1968} and \citet{Buchler1981} applied non linear processes 
to the physics of pulsating variable stars \citep[see also][]{Regev1981, Buchler1993}. 

Researches on the instabilities in accretion discs started in the seventies 
\citep[e.g.][]{Lightman1974, Pringle1973, Shakura1976} 
and up to now a large number of papers was produced.
\citet{Taam1984}, in particular, using numerical integration of
the non-linear disc equations and applying the $\alpha$ prescription for the viscosity
\citep{Shakura1973}, investigated a thermal-viscous instability and obtained 
a few theoretical light curves having recurrent spikes.
After the discovery of the $\rho$ class variability in \grs, \citet{Taam1997}
analysed the time and spectral properties of the bursts and proposed an 
interpretation based on the instability discussed in the previous paper.
The evolution of thermal viscous instabilities in an accretion disc is associated 
with a limit cycle \citep[e.g.][]{Szuszkiewicz1998}, generally described by means 
of S-shaped equilibrium curves in a plot of temperature or accretion rate vs disc 
surface density, where the different signs of the slope correspond to stable and
unstable equilibrium states \citep[e.g.][]{Abramowicz1995}.
Examples are the models developed by \citet{Watarai2001} \citep[see also][]{Mineshige2005}
for a slim-disc \citep[][]{Abramowicz1988}
as indicated by the energy spectral results for \grss of \citet{Vierdayanti2010} 
and \citet{Mineo2012}, and by \citet{Janiuk2000, Janiuk2002}, who included dissipation
processes due to the presence of a corona and an outflow.
More recently \citet{Potter2017} demonstrated that a similar 
profile is found for a thermal-viscous instability induced by a turbulent
accretion disc stress factor depending upon the magnetic Prandl number.

It is interesting that the topology of one of the 
equilibrium curves of the proposed ODE system presents a S-shaped pattern 
similar to those derived from numerical solutions
of instability disc equations.
We will show that this allows for both stable and unstable equilibrium 
states which provide a useful and accurate modelling of the observed X-ray light 
curves, including many details.
In this first paper, we limit our analysis to the case of a steady input function, 
while the solutions for a variable input and the extension to other classes will be 
discussed in the companion paper \citet[][hereafter Paper II]{Massaro2020}, 
together with a tentative interpretation of the equations on the basis of literature
disc instability calculations.

\section{The MHR non-linear ODE system}
\label{sct:MHR}

To reproduce the rich and complex behaviour of \grs, we considered a non-linear 
system of ODE as those used for describing quiescent and bursting signals in 
neuronal arrays.
This approach offers the possibility of describing transitions between stable and 
unstable equilibrium states with the onset of limit cycles.
Mathematical aspects of this important topic were deeply investigated in the
past half century and an extremely wide and technical literature is available
(see, for instance the textbooks of \citealt{Izhikevich2006} or 
\citealt{Gerstner2014}).

The general formulation of the Hindmarsh-Rose (hereafter HR) system considers 
three ODE for the dynamical variables $x$, $y$ and $z$, involving changes 
on different time scales; it is generally written as:

\begin{eqnarray}
\label{eq1}
\frac{dx}{dt} &=&  \frac{1}{A}[ P_3(x) + b_1 y - z] \nonumber \\
\frac{dy}{dt} &=& P_2(x) - b_2 y          \\      
\frac{dz}{dt} &=& \varepsilon [ s (x -x_0) - z ]  \nonumber 
\end{eqnarray}

\noindent
where $A$ is a scale factor and $P_2(x)$ and $P_3(x)$ are two polynomials of 
second and third degree, respectively:

\begin{eqnarray}
P_3(x) &=& -a_1 x^3 + a_2 x^2 + a_3 x + a_4  \nonumber \\
P_2(x) &=& -a_5 x^2 + a_6 x + a_7 
\label{eq2}
\end{eqnarray}

\noindent
With the parameters' values $A = 1$, $a_1 = 1$, $a_2 = 3$, $a_3 = a_6 = 0$, 
$a_4 = 5.0$, $a_5 = 5$, $a_7 = -3$, and $b_1 = b_2 = 1$, Eqs.~\ref{eq1} and 
\ref{eq2} give the classic HR model (\citealt{Hindmarsh1984,  Hindmarsh2005} 
and the tutorial paper by \citealt{Shilnikov2008}).
In our work we adopted a modified system of the HR equations and substituted the 
variable $z$ with an external input function of the time $J(t)$.
The system includes, therefore, only two equations that, adopting a notation 
similar to that already used in \citet{Massaro2014}, are written as:

\begin{eqnarray}
\frac{dx}{dt} &=& -\rho x^3 + \beta_1 x^2 + y + J(t) \nonumber \\
\frac{dy}{dt} &=& -\beta_2 x^2 - y  
\label{eq3}
\end{eqnarray}

\noindent
where we indicate the two variables with $x$ and $y$, as before, and the
signs of the various terms were taken to have the parameters' values positive.

This ODE system can be easily written in a single equation in the two variables:

\begin{equation}
\frac{dx}{dt} = - \rho x^3 - (\beta_1/\beta_2) \frac{dy}{dt} + (1 - \beta_1/\beta_2) y + J(t) 
\label{eq4}
\end{equation}

\noindent
or the equivalent one:

\begin{equation}
\frac{dx}{dt} = - \rho x^3 + (\beta_1 - \beta_2) x^2 - \frac{dy}{dt} + J(t) 
\label{eq4m}
\end{equation}

\noindent
that makes clear the role of the variable $y$ and its derivative in the evolution of
$x$ and that when the cubic term turns to be the dominant one it would be responsible
of the fast decrease just after the maximum.

In our calculations the solutions for the $x$ variable correspond to those for 
the X-ray luminosity, where there is the largest energy release, while the $y$ plays 
the role of a state variable of the disc plasma directly related to the radiative 
energy dissipation.
Moreover, the latter quantity must be relevant in the developing of unstable processes
and of conseguent limit cycles.
We will discuss this subject in \citetalias{Massaro2020}.

As demonstrated inq the Appendix ~\ref{appendix1}, the system of Eq.~\ref{eq3} 
is also equivalent to the single differential Eq.~\ref{eq3:app1} for $x(t)$, and $y(t)$ 
plays the role of a separation variable to split it into two equations.
The choice of $y(t)$ is somewhat arbitrary and usually it is defined to obtain
a system well suited for studying the stability of equilibrium points.
An example of an equivalent ODE system with another variable instead of $y$
is also given in the Appendix~\ref{appendix1}.
Note also that  Eq.~\ref{eq3:app1} contains the term $x^3$ and not a linear one
as for a harmonic oscillator.
In \citet{Massaro2014}, the considered ODE system included two linear terms instead 
of the two quadratic ones; the solution for $y(t)$ 
resulted very similar to the curve of the mean photon energy, but this finding 
does not apply to the system considered in the present work.

The system in Eq.~\ref{eq3} contains three free parameters and an input function.
The parameter $\rho$ is unimportant because it can be eliminated by means of the 
transformations
$\tilde{x} = \sqrt{\rho} x$, $\tilde{y} = \sqrt{\rho} y$, 
$\tilde{J}(t) = \sqrt{\rho} J(t)$, and dividing the two other parameters 
by $\sqrt{\rho}$, as it is easy to verify.
We apply here the equivalent choice of fixing $\rho = 1.0$, without any loss
of generality, and adopt the simplifying assumption 
$\beta_1 = \beta_2 = \beta > 0$: 

\begin{eqnarray}
\frac{dx}{dt} &=& - x^3 + \beta~ x^2 + y + J(t) \nonumber \\
\frac{dy}{dt} &=& - \beta~ x^2 - y  
\label{eq5}
\end{eqnarray}
 
\noindent
that gives solutions which reproduce the majority of classes and can be used for 
describing the equilibrium states and the transitions between them.
Note that in this case the linear $y$ term is cancelled in Eq.~\ref{eq4}.
These assumptions will be {\it relaxed} in \citetalias{Massaro2020} to extend the 
model to other variability classes.
In the following we will refer to these equations as Modified Hindmarsh-Rose system
(hereafter MHR).
Numerical computations were performed by means of a Runge-Kutta fourth order 
integration routine \citep{Press2007}.

\section{Nullclines, equilibrium points and stability}
\label{sct:nullc}

In the simple case of a constant $J(t) = J_0$ the equilibrium conditions for the 
system of Eq.~\ref{eq5}, i.e. $\dot{x} = \dot{y} = 0$, are

\begin{eqnarray}
 y &=&  x^3 - \beta~ x^2 - J_0 \nonumber \\
 y &=& - \beta~ x^2  
\label{eq6}
\end{eqnarray}

\begin{figure}
\includegraphics[height=7.9cm,angle=-90,scale=1.0]{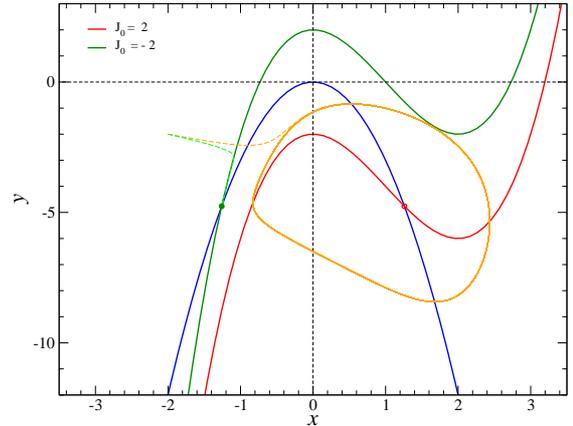}
\caption[]{
{\it Nullclines} for the system of Eq.~\ref{eq5} with $\beta = 3$ and for  $J_0$ 
values equal to $-2$ (green curve) and $2$ (red curve); the blue curve is the parabola 
given by the second formula of Eq.~\ref{eq6}.
Two phase trajectories (dashed lines) with the same initial conditions and 
corresponding to the two $J_0$ values are also shown: the green one moves along its
{\it nullcline} to the stable equilibrium point (green filled circle) while the orange 
line describes a closed orbit around the unstable equilibrium point (red open circle).
}
\label{f1ncl}
\end{figure}

\noindent
This system admits only the real solution $x_* = J_0^{1/3}$, $y_* = - \beta ~J_0^{2/3}$.
In Fig.~\ref{f1ncl} we plotted in the plane $x,y$ the curves of Eq.~\ref{eq6}, named 
{\it nullclines}, which intersect at the equilibrium point, that results always stable 
for $J_0 < 0$ while it has an unstable interval for $J_0 > 0$, as explained in 
Appendix~\ref{appendix2}. 

\begin{figure}
\includegraphics[height=8.0cm,angle=-90,scale=1.0]{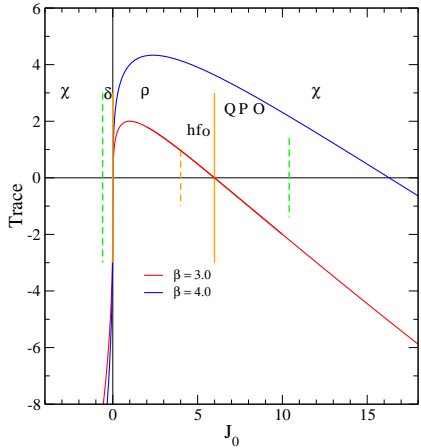}
\caption[]{
The plot of the traces of the Jacobian for two values of $\beta$ as function 
of the mean input $J_0$.
Stable  intervals ($\chi$, QPO) for the curve relative to $\beta=3.0$ are outside 
the vertical orange segments, and the unstable interval ($\rho$) is between them.
Dashed lines within the stable and unstable regions mark the intervals in 
which other variability classes and behaviours are obtained as explicitly 
indicated in the graph.
}
\label{f2tr}
\end{figure}

This stability analysis shows that it depends upon the sign of the trace $Tr$
of the Jacobian of the system evaluated at $(x_*, y_*)$ (see Eq.~\ref{eqA2-5}).
Fig.~\ref{f2tr} shows the plots of $Tr$ as function of $J_0$ for two values of 
$\beta$, 3.0 and 4.0; the two vertical orange lines 
correspond to the zeroes of the trace and delimit the unstable interval 
for $\beta = 3.0$, that, using the formulae given in Appendix~\ref{appendix2}, 
results [0.0061792, 5.99382].
It is important to note that the instability interval depends only upon $\beta$ 
but not upon $J_0$; thus a change of this parameter moves the location of the
equilibrium point allowing transitions between stable and unstable states.
The above limits of the $J_0$ interval define the states at which the transition 
from stable to unstable equilibrium occurs, and therefore they rule the onset 
or the disappearance of the spiking behaviour.
Examples of stable and unstable dynamical solutions are illustrated in Fig.\ref{f1ncl}, 
where two trajectories corresponding to the values of $(x(t)$, $y(t))$ of the 
system in Eq.~\ref{eq5} are also plotted:
they start from the same initial position, $x_0 = -2.0$, $y_0 = -2.0$, but, while 
the one for $J_0 = -2.0$ reaches the green {\it nullcline} and then moves toward 
the corresponding equilibrium point, the other (orange trajectory) crosses the 
blue {\it nullcline} for $J_0 = 2.0$ and evolves to a 
closed orbit (limit cycle) around the unstable (red) equilibrium point
(see Sect.~\ref{sect:class_rho}).

It is interesting to note that the unstable interval for $\beta = 3.0$ is 
[0.1835, 1.8165] and it is entirely contained in the interval [0, 2.0] 
corresponding to the portion of the $x$ {\it nullclines} with a negative slope,  
as it is easy to verify from the roots of the $x$ derivative of the first of 
Eq.~\ref{eq6}.

\section{Synthetic light curves}

We show in this Section how the numerical solutions of the simplified system of 
Eq.~\ref{eq5} reproduce the main features of several variability classes and, 
particularly, four of them are obtained with a constant input level.
Light curves are obtained by considering only the solutions for $x(t)$, which
can be scaled by means of a linear transformation to the X-ray photon flux from
the disc that is the main contributor to the luminosity in this band.
Solutions for the other variable $y(t)$ will not be considered in the present 
paper and, in general, they could not directly be related to another observable
quantity.
However, one of the results, shortly presented in the Sect.~\ref{ssect:rhoy}, 
exhibits a high similarity with the disk inner radius as measured by 
\citet{Neilsen2012}.

For some other classes, it is necessary to assume an input function variable on 
time scales similar to those detected in \grss light curves, with the exception 
of the spiking that is originated by non-linear instabilities.
We will describe the analysis of the classes requiring a variable input in 
\citetalias{Massaro2020}.

In this work, we limit our analysis to an input function $J(t)$ having
the simple form:

\begin{equation}
   J(t) = J_0 + C~ r
\label{eq8}
\end{equation}

\noindent
where $J_0$ is a constant component and $r$ is a random number with a uniform 
distribution in the interval [$-0.5$, $0.5$], $C$ the constant amplitude of 
random fluctuations and the resulting standard deviation $\sigma_J = C/(2 \sqrt{3})$.
The quantity $r$ is useful to simulate statistical fluctuations, as for instance 
those expected from turbulences in the disc, and to make our numerical results 
more similar to the observed ones.
Of course, more realistic models would require the knowledge of the turbulence 
spectrum, that is not yet determined.
The physical interpretation of $J_0$ is not apparent in this formulation:
in \citet{Massaro2014}, we assumed that it may be related to the local mass 
accretion rate $J(\dot{m})$, but this hypothesis will be newly reconsidered in 
the present and in \citetalias{Massaro2020} on the basis of the MHR results. 

\begin{figure}
\includegraphics[height=7.9cm,angle=-90,scale=1.0]{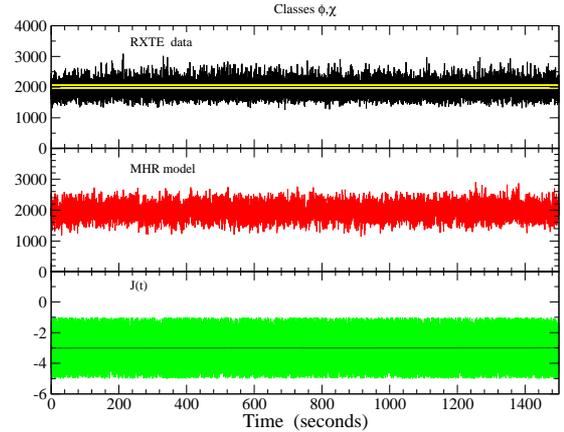}
\caption[]{
{\it Top}: segment of a RXTE/PCA light curve of the $\chi$ class (ID 10408-01-22-02)
in the 2-12 keV energy interval, the horizontal yellow belt around the mean level is 
the mean amplitude expected from Poisson noise. 
{\it Centre}: model light curves obtained by a numerical solution of Eq.~\ref{eq5} 
{using as input function the curve in the bottom panel}.   
{\it Bottom}: input function consisting of the superposition of a constant level 
$J_0 = -$3.0, equal to the mean value represented by the dark green horizontal line, 
and a white random noise given by $C = 4.0$.
}
\label{fchi}
\end{figure}

The other parameter to be determined is $\beta$.
Considering that we used a linear transformation to scale our numerical results 
to the actual time and amplitude of the light curves, we decided to use parameters'
value in a range between 0 and 10, to manipulate small numbers which reduce the 
possibility of numerical troubles in the numerical integration.
From our results, we found that a $\beta$ value between 3.0 and 4.0 gives 
solutions in good agreement with the data: the higher value producing spikes 
slightly narrower than those obtained with the lower one while no apparent 
difference is found for the stable light curves.
In the following we will present the results obtained assuming $\beta = 3.0$ 
because transitions between different variability classes are obtained for $J_0$ 
changes within a rather narrow range, as explained in Appendix~\ref{appendix2}.
The values of this parameter used for computing the following light curves are 
given in Table~\ref{tab:param}.
However, in \citetalias{Massaro2020} we will show that a few classes of light 
curves are better reproduced for different choices of this parameter.
The amplitude of the random fluctuating component was generally fixed to 
 $C = 4.0$, while the value 3.5  was considered in Sect.~\ref{sect:qpo} for 
investigating the QPO origin.

For the comparison of MHR results with data we used several RXTE/PCA observations 
selected as examples of the considered classes.  
The  light curves are accumulated with standard procedures in the energy range 
2-12 keV.
 
\subsection{Classes $\phi$, $\chi$}

Light curves of these two classes are characterised by a rather stable or slowly 
variable mean level with fluctuations typical of a statistical noise.
The main difference between these two classes is in the hardness ratio and this 
changes are not taken into account in our mathematical modelling that is limited 
to reproduce the observed patterns of the X-ray light curves in the energy range 
2-12 keV, dominated by the disc emission. 
As an example, an observed light curve relative to the class $\chi$ is shown in 
the top panel of  Fig.~\ref{fchi}.
Note that fluctuations in the light curve are much higher than the Poissonian noise 
associated with the observed counts, whose mean amplitude is represented by the 
yellow strip around the mean count level.
These fluctuations could be originated by fast random changes intrinsic to the 
disc as one can expect from plasma turbulence or other similar stochastic processes.

In Fig.~\ref{fchi}, the observed light curve is compared with the output of the 
MHR model of Eq.~\ref{eq5} for $J_0 = -3.0$ and the corresponding input function 
$J(t)$ is represented by the green data series in the bottom panel where the darker 
line indicates the mean value.
Fluctuations in the computed values are clearly due to the random white noise and, 
for $C = 0$, they are absent and a constant equilibrium solution is obtained.

We note that the MHR model reproduces the $\phi$ and $\chi$ classes when 
$J_0$ is increased up to values high enough that the equilibrium point is again 
in a stable region ($J_0>6$), however this transition from the unstable region to 
the stable one implies the disappearance of the limit cycle and the onset of a 
QPO phenomenon as shown in Sect.~\ref{sect:qpo}.

\begin{figure}
\includegraphics[height=7.9cm,angle=-90,scale=1.0]{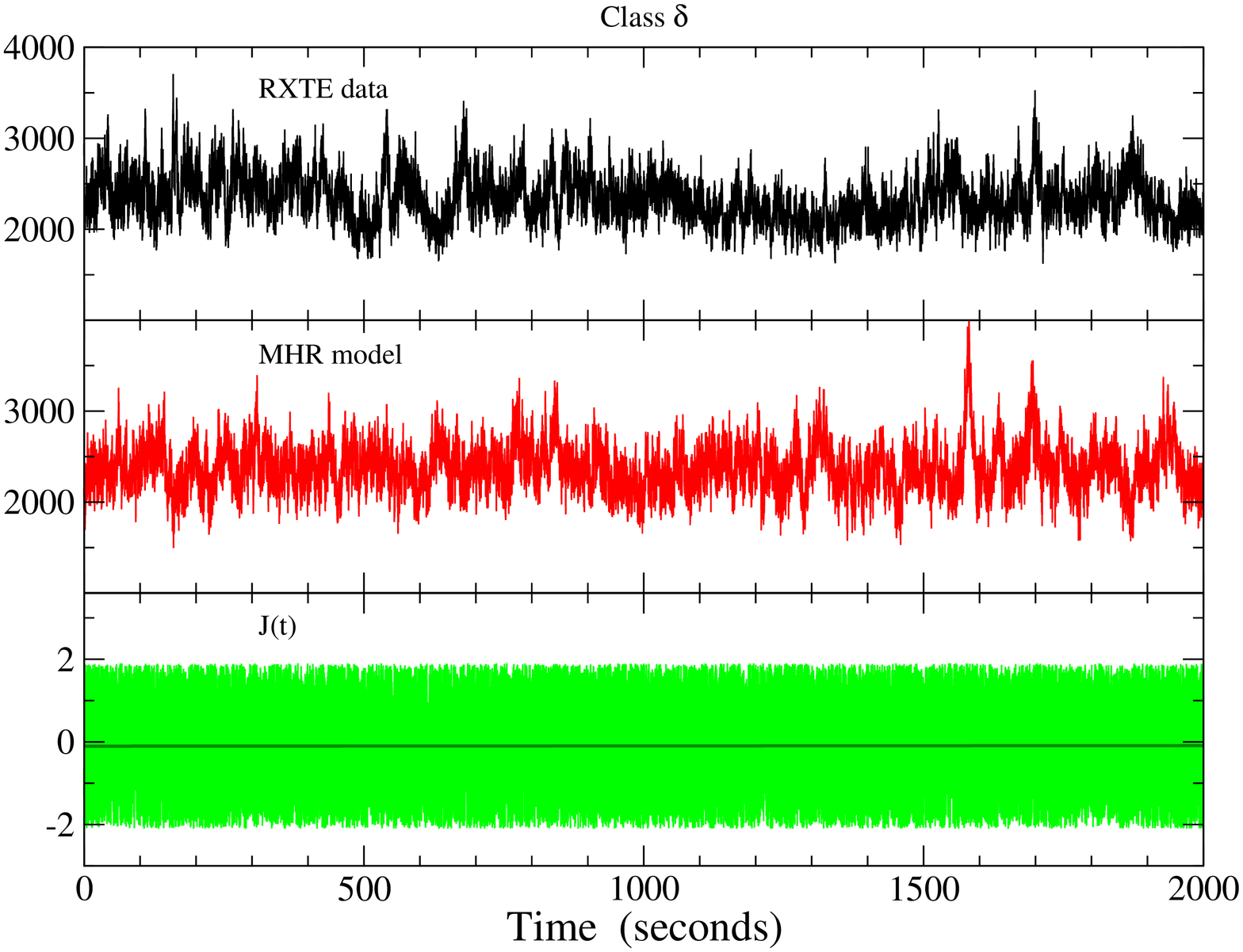} \\
\includegraphics[height=7.9cm,angle=-90,scale=1.0]{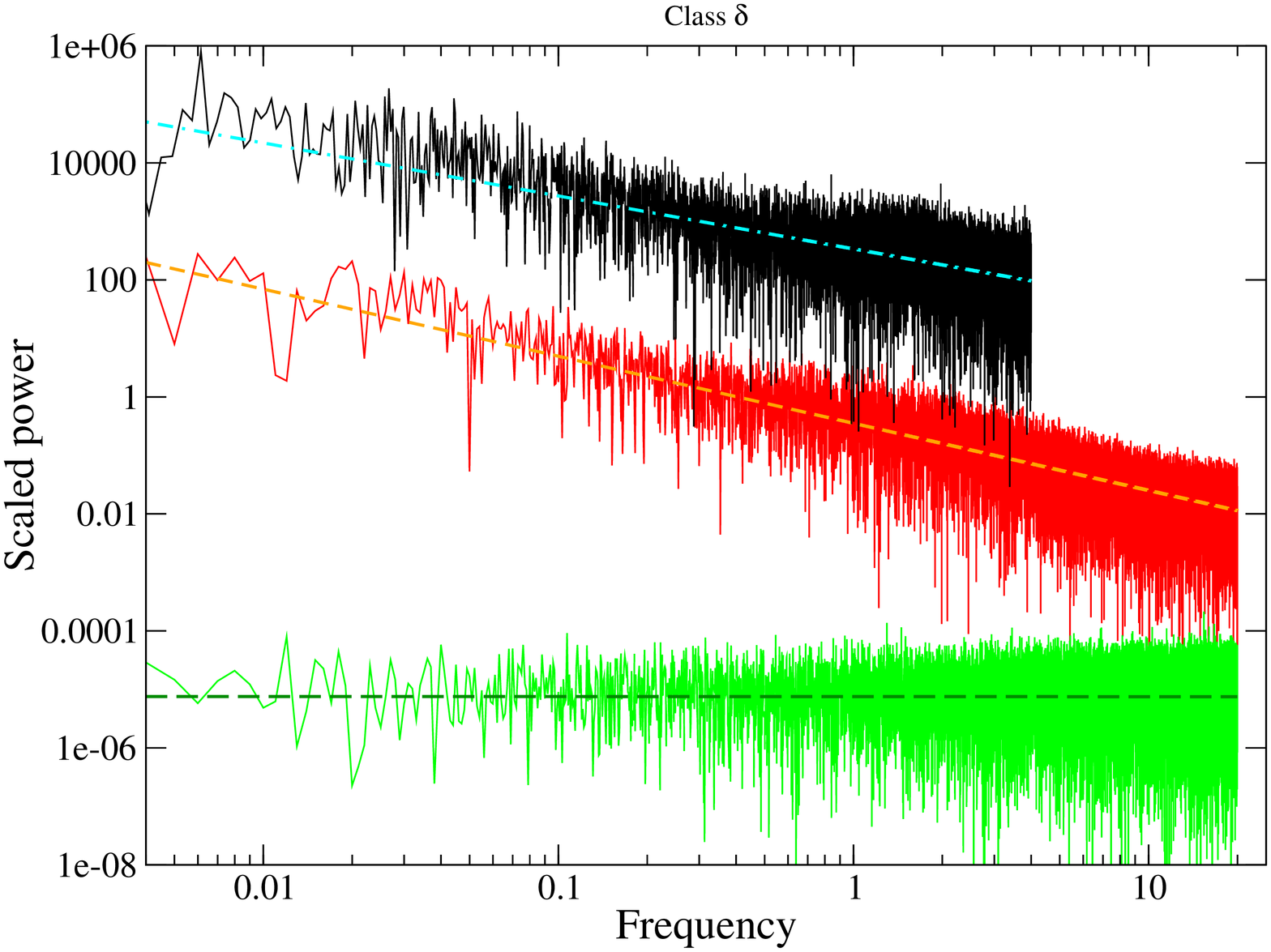}
\caption[]{
{\it Upper plot}: segment of a RXTE/PCA light curve of the $\delta$ class 
(ID 10408-01-13-00) in the 2-12 keV energy interval (black curve in the top panel);
light curve obtained by means of the MHR model (red curve in the central panel)
using as input function the green curve in the bottom panel obtained by the 
superposition of a white random noise $C = 4.0$ to a constant level $J_0 = -0.1$; 
the mean value is represented by the dark green horizontal line.  \\
{\it Lower plot}: power density spectra of the $\delta$ light curves in the 
upper plot, for the observed data (black), for the curve computed with the MHR 
model (red) and for the input $J(t)$ (green).
The dashed cyan, orange and dark green lines are the relative power law best fits.
}
\label{fdelta}
\end{figure}

\subsection{Class $\delta$}

According to \citet{Belloni2000}, light curves of the $\delta$ class 
appears to have a rather stable mean level but with a red noise-like variability.  
Very similar results are obtained by means of the MHR model when the 
$J_0$ value increases approaching to 0.
RXTE data and a computed light curve are shown in top and centre sections of
the upper panel in Fig.~\ref{fdelta}, respectively.
We can use these curves to apply the Fourier analysis and evaluate the 
Power Density Spectrum (PDS) that indeed presents a red noise with a power law 
distribution having the exponent equal to $-1.1$, remarkably close to the one
of the true data that is $-0.91$,
while that of the input $J(t)$ corresponds to a flat white noise (Fig.~\ref{fdelta}, 
lower panel).
We can therefore conclude that the non linearity of the MHR model also acts as a 
red noise generator, when the equilibrium points approaches the instability region.
Also for this class the assumption $C \ne 0$ is necessary to obtain these patterns.
Eliminating the noise the MHR solution is a constant equilibrium value which changes 
to a small amplitude oscillation and to the spiking behaviour for increasing $J_0$.

As shown in the following subsection a further increase of $J_0$ will
produce $\rho$ class light curves.
The $\delta$ class should therefore be considered as a transition class between the 
$\chi$ and $\rho$ classes. 

\begin{table}
\caption{Parameters' values adopted in the numerical calculations of the light curves 
for the variability classes of \grs.}
{\small
\begin{tabular}{crrr}
\hline
   Class     & $\beta$ &  $C$  &  $J_0$  \\
\hline
             &        &       &         \\ 
 $\chi$      &   3.0  &  4.00 &  $-$3.0 \\ 
 $\delta$    &   3.0  &  4.00 &  $-$0.1 \\ 
 $\rho_d$    &   3.0  &  4.00 &    0.05 \\
 $\rho$      &   3.0  &  4.00 &    0.80 \\ 
\hline

\end{tabular}
}
\label{tab:param}
\end{table}

\subsection{Classes $\rho$, $\rho_{d}$}
\label{sect:class_rho}

\begin{figure}
\includegraphics[height=7.9cm,angle=-90,scale=1.0]{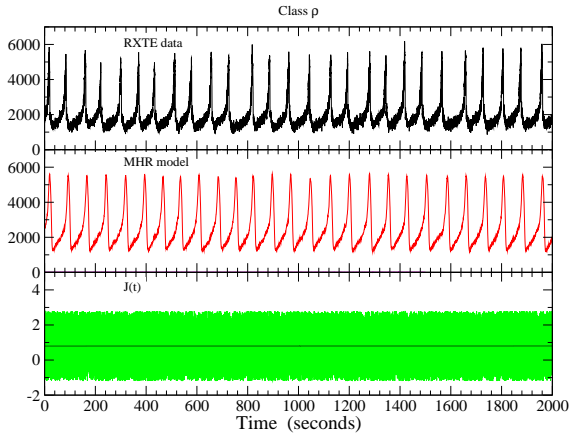}
\caption[]{
{\it Top}: segment of a RXTE/PCA light curve of the $\rho$ class observation 
ID 20402-01-31-00 in the 2-12 keV energy interval;
{\it Central}: light curve obtained with the MHR model using the input curve 
plotted in the bottom panel;
{\it Bottom}:  input function $J(t)$, with random fluctuations obtained with 
$C = 4.0$ superposed to a constant value $J_0 = 0.8$; its mean value is the dark 
green line.
}
\label{frho}
\end{figure}

\begin{figure}
\includegraphics[height=7.9cm,angle=-90,scale=1.0]{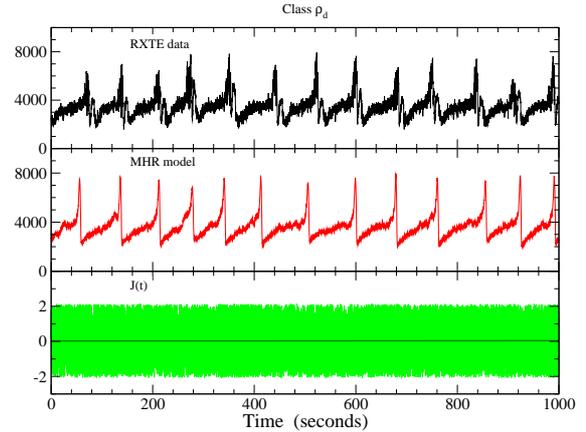}
\caption[]{
{\it Top}: segment of a RXTE/PCA light curve of the $\rho_d$ class of the 
observation ID 10408-01-41-00 in the 2-12 keV energy interval.
{\it Central}: light curve (red) obtained with the MHR model using the input 
model in the bottom panel.
{\it Bottom}: input function $J(t)$ with random fluctuations obtained with 
$C = 4.0$ superposed to a constant value  $J_0 = 0.05$; its mean value is the 
dark green line.
}
\label{frhod}
\end{figure}

The class $\rho$ is likely the most interesting and the most studied one: it consists
of nearly regular series of spikes with a recurrence time variable in the range from
about 40 s to more than 100 s; the shape of the spikes is characterised by an initial  
rather slow rise followed by a faster increase and a similarly fast decline 
(see Fig.~\ref{frho}).
Moreover, in several cases the spikes have a double or even a multiple structure.
 \citet{Massaro2010} introduced a multiplicity parameter $p$ to number the peaks 
apparent in the fine structure and \citet{Yan2017}\citet{Yan2017} defined the 
`subclasses' $\rho_1$ and $\rho_2$ corresponding to $p$ equal to 1 or 2, which are 
the most frequent patterns.
We here introduce another subclass $\rho_d$, which has spikes similar to those of the 
typical $\rho$ one, but with the occurrence of a dip just at the their end, frequently 
followed by a fast rise and a plateau preceding a new fast rise of the following peak 
(see Fig.~\ref{frhod}).
The $\rho_d$ mean recurrence time is usually longer than that of $\rho$ spikes.
This subclass was the first one observed by \citet{Taam1997}, who 
reproduced its main features by computing the light curve due to the disc instability 
previously investigated by \citet{Taam1984}.

Our mathematical model is able to produce a spiking behaviour when the mean $J(t)$ 
is further increased above the stability threshold: a transition from stable to  
unstable equilibrium occurs and a limit cycle is established.
A similar behaviour was also found for the FitzHugh-Nagumo model analysed in  
\citet{Massaro2014} and \citet{Ardito2017}.
Fig.~\ref{frho} and Fig.~\ref{frhod} (middle panel)  show the spiking behaviour of 
the main class and of the subclass that are remarkably similar to the observed ones.
The variable recurrence time of spikes, clearly apparent in the data panel in 
Fig.~\ref{frho} (top panel), can be due local changes of the mean level of $J(t)$, 
as it will be discussed in detail in the following section.
Peaks in the $\rho_d$ data have generally a double structure, as already noticed by 
\citet{Taam1997} and in other observations the $\rho$ peaks exhibit also a more
complex fine structure  \citep[see, for instance,][]{Belloni2000, Massaro2010}.
Here we modelled only a single peak pattern and these more structured profiles require
some other assumption on the input function.
Note also that this light curve is more irregular than the one of the $\rho$ class;
moreover, its $J_0$ is slightly lower and therefore it appears as a 
transition between the $\delta$ and the $\rho$ classes.

\begin{figure}
\includegraphics[height=7.9cm,angle=-90,scale=1.0]{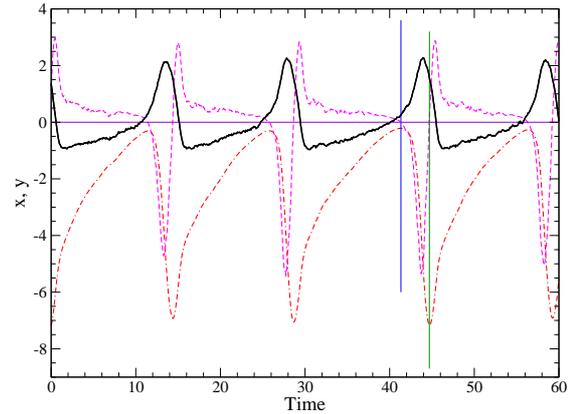}
\caption[]{
Numerical solutions of the variable $y(t)$ (red curve) and its first derivative
(magenta curve) compared with the $x(t)$ (black curve) when in the $\rho$ spiking 
class.
The blue and green vertical lines limit the interval between a maximum and a 
minimum of $y(t)$ where it is very quickly decreasing that corresponds to the 
highest emission in the $x(t)$ peak.
The violet horizontal line marks the zero level. }
\label{rho-xy}
\end{figure}

\subsubsection{The $y$ variable}
\label{ssect:rhoy}

For a better understanding of the solutions for the $\rho$ class it is useful to
consider the behaviour of the $y$ variable and of its time derivative.
With the adopted parameters' values, Eq.~\ref{eq4} becomes:

\begin{equation}
\frac{dx}{dt} = - x^3 - \frac{dy}{dt} + J(t) 
\label{eq9}
\end{equation}
\noindent
that shows the relevant role of the $y$ derivative in the evolution of the burst
profile.

\begin{figure}
\includegraphics[height=7.9cm,angle=-90,scale=1.0]{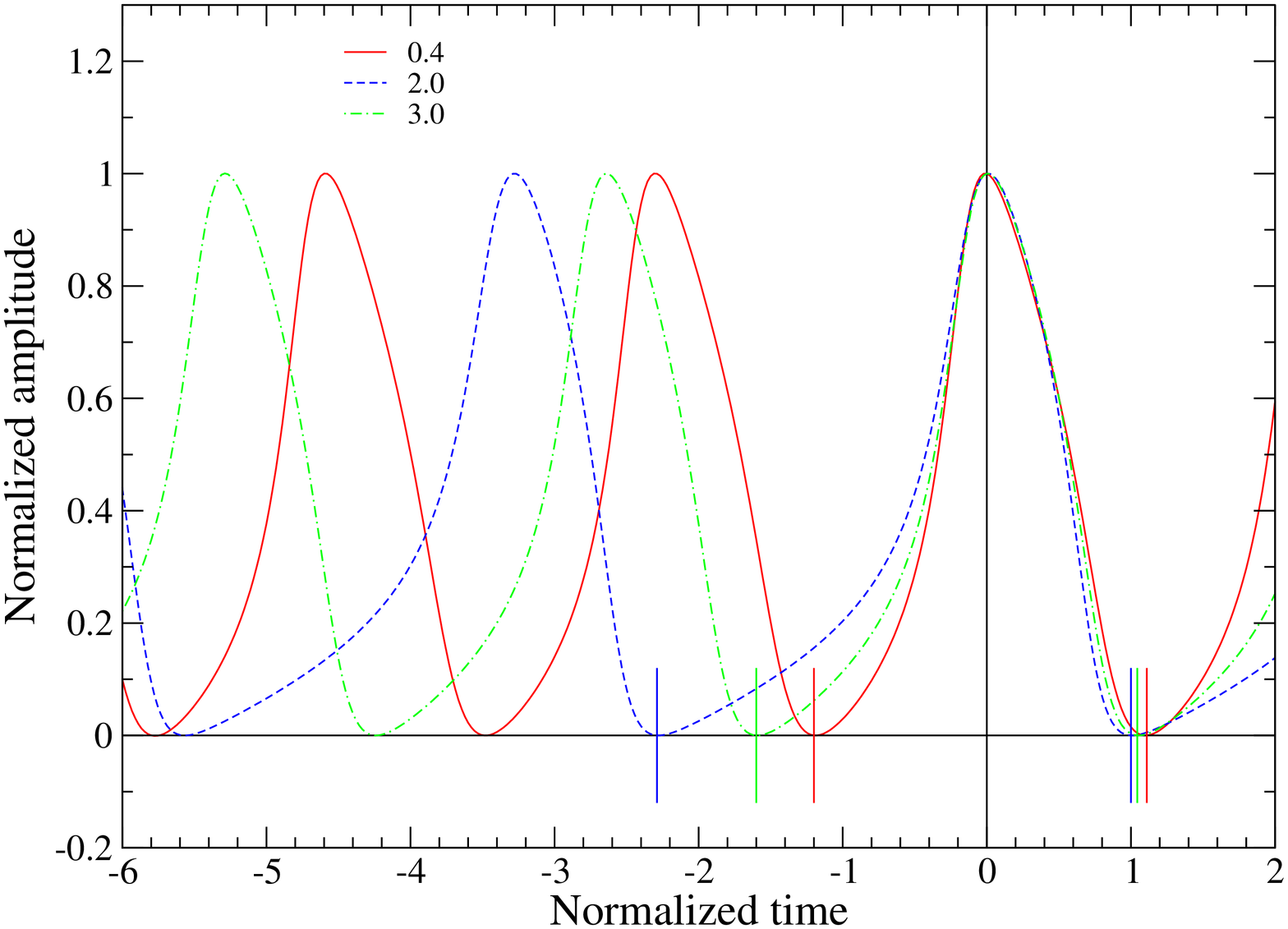}
\includegraphics[height=7.9cm,angle=-90,scale=1.0]{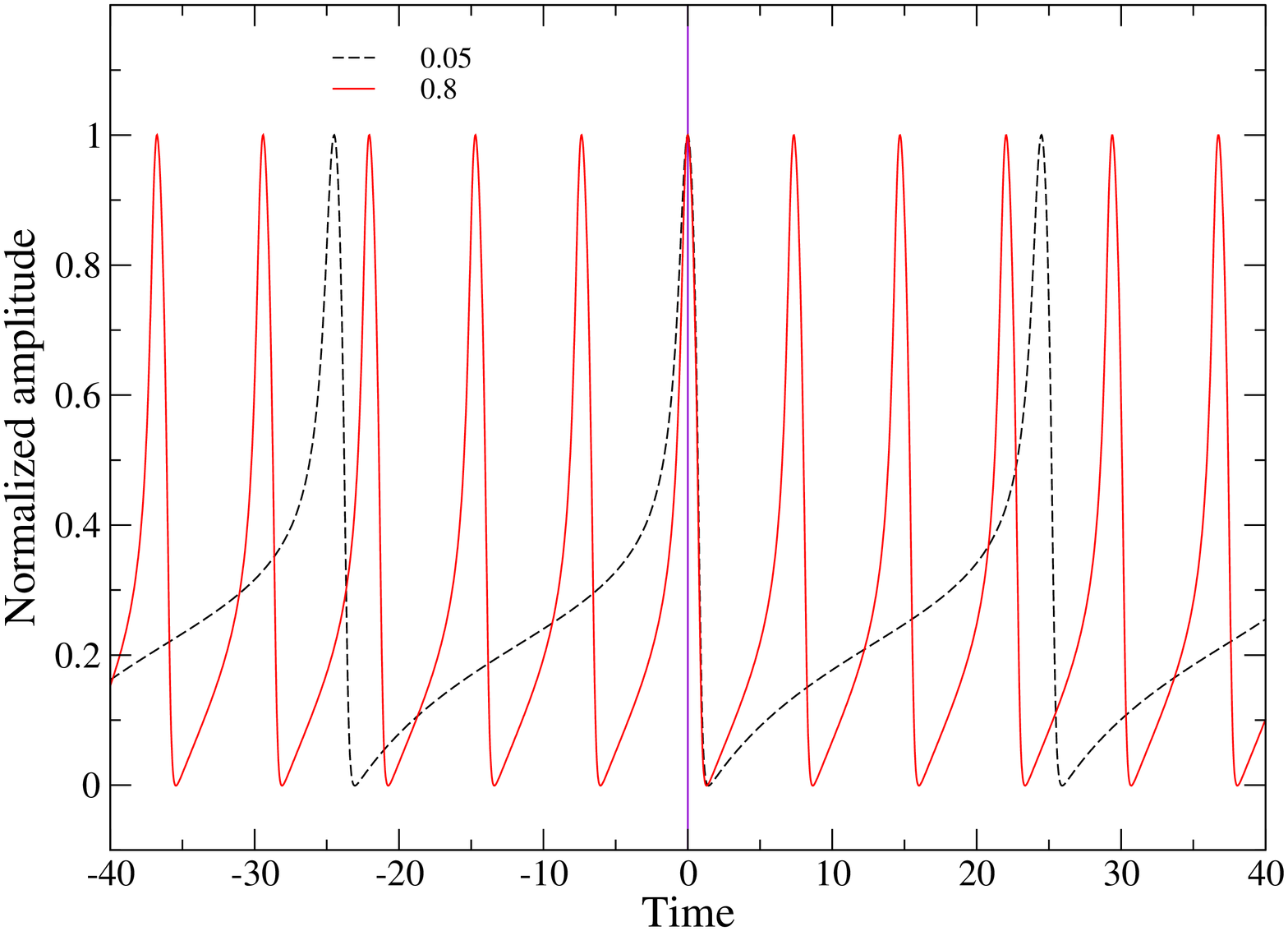}
\caption[]{{\it Upper plot}:
Normalized profiles of three spikes of the $\rho$ class obtained by 
the MHR model for three values of $J_0$ without the random component.
Vertical bars mark the initial and final points of the corresponding colour 
spikes.
Curves, after the subtraction of the minimum level, are
normalized in amplitude by taking the maximum value equal to unit, and
in time by fixing to 1.0  the duration of the interval between the maximum 
and the end point of the spike for $J_0 = 2.0$  (blue line). \\
{\it Lower plot}: comparison between results for a typical $\rho$ (red) and 
a $\rho_d$ (black) profile obtained for a $J_0$ close to the stability boundary.
}
\label{frhoprof}
\end{figure}

Fig.~\ref{rho-xy} shows the numerical solutions for $y(t)$ and its derivative 
corresponding to the $\rho$ bursting sequence of  $x$.
In this case, the variables' values are not scaled to the measured count rates 
and no constant offset was added, as in Fig.~\ref{f1ncl}.
The profile of $y$ consists of an initially fast increase followed by a milder 
hump up to the maximum and then by a sudden and very fast decrease.
This shape is very similar to the one found by \citet{Neilsen2012} (see their 
Fig. 6) for the inner radius of the accretion disc in the time resolved spectral 
analysis of the $RXTE$ observation 40703-01-07-00, performed using the model 
developed by \citet[][{\sc ezdiskbb} in XSPEC ]{Zimmerman2005}.
The two vertical lines in Fig.~\ref{rho-xy} limit one of the decreasing intervals,
where the $y$ derivative is negative, to show that it corresponds to the interval
of the peak in the $x$ curve.
Again, there is a similar phase shift between the maxima of the two variables
as in those of \citet{Neilsen2012}.
We recall that, in the emission models, the luminosity is proportional to the square 
of the inner radius and therefore it can exhibit variations with the same time
scales.
This correspondence supports the association of the $y$ with a disc quantity
which follows a limit cycle as the luminosity.
In the discussion of \citetalias{Massaro2020} we will present a possible 
interpretation to be verified by means of calculations of thermal-viscous
instabilities.

\section{Structure and period of the $\rho$ class spikes}
\label{sect:rho}

The two qualifying observable quantities of the $\rho$ class spiking are
the profile and the recurrence time of spikes.
Several observations \citep[see, for instance,][] {Neilsen2011,Weng2018} have
shown that the spike structure is highly variable and that it frequently can
exhibit two or more peaks.
However, in many occasions and over time intervals of several thousands of 
seconds, the mean profile appears to be stable  \citep{Massaro2010}.
The solutions of the MHR model, whitout any random component in $J(t)$, give 
smooth and stable profiles with a duration, and consequently the recurrence 
time, depending upon $J_0$.
Three profiles, computed for $J_0$ in the range [0.4, 4.0], are shown in
the upper plot of Fig.~\ref{frhoprof}.
For a better comparison of these profiles we subtracted the constant offset 
level and normalized the maxima to unity. 
Moreover, the time distance between the maximum and the end of
the spike was fixed also to unity for the longest one ($J_0 = 0.4$,
red curve), and  then we used this scale factor for reducing  
the periods of the other two profiles which where aligned at the
time of the maximum.
One can see that the change of the spike duration is mainly due to the 
rising section, while the decaying part remains more stable, slightly increasing
with $J_0$.
It follows that spikes evolve to be more symmetric when the recurrence time 
decreases.
In  practice, one can consider that for $J_0 \gtrsim 4.0$ the spiking
behaviour changes to a nearly sinusoidal high frequency oscillation 
(hereafter `hfo') whose frequency changes very little for further increases 
of $J_0$.

\begin{figure}
\includegraphics[height=7.9cm,angle=-90,scale=1.0]{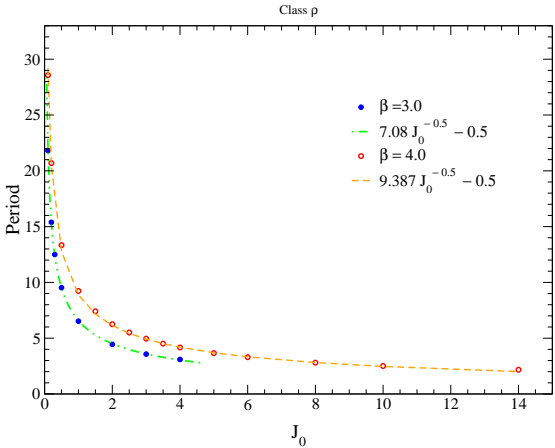}
\caption[]{
The period of spikes in the $\rho$ class for two values of $\beta$ as 
function of $J_0$.
Blue and red circles are the results of numerical calculations, while
the dashed lines are the best fit interpolations with a fixed exponent.
}
\label{frhoper}
\end{figure}

It is also interesting to see how spike profiles change when the values of
$J_0$ are just above the stability boundary and produce the $\rho_d$ pattern.
This comparison is shown in the lower panel of Fig.~\ref{frhoprof}, where two 
normalised MHR results are plotted.
Again major differences are in the rising segment of the spike: in the 
$\rho_d$ case it starts with a high slope that decreases and turns to increase 
again to reach the maximum.
This particular shape can be affected by fluctuations which can amplify
the slope changes moving the input value across the stability threshold,
thus producing the dips at the spikes' end, which are thus explained by the
non-linear dynamics of disc oscillations.

The changes of the recurrence time, essentially due to the variable duration 
of the rising part of spikes, are thus explained by the occurrence of a {\it slow}
and a {\it fast} time scale, as occurs in the limit cycles of non-linear 
oscillators.

A very interesting property of the $\rho$ class spikes is that their periodicity 
depends on the value of $J_0$.
To obtain a functional relationship we computed several light curves, without 
the random component (i.e. fixing $C = 0.$ in Eq.~\ref{eq8}), for $J_0$ values 
ranging from 0.1 to 4.0 and two values of $\beta$.
The spike period was then evaluated by means of a Fourier periodogram and the 
resulting values are plotted in Fig.~\ref{frhoper}.
A very well defined and regular decreasing trend is clearly apparent that is
described either by a simple power law:

\begin{equation}
   P(J_0) = K(\beta)~J_0^{-s} 
\label{eq10}
\end{equation}

\noindent
For both $\beta$ values the resulting exponent was $s = 0.535$, very close and 
only slightly different than 0.5.
We then fixed the exponent to this value and included an additional constant term 
in the best fitting formula:

\begin{equation}
   P(J_0) = K(\beta)~ J_0^{-0.5} + q   ~~~~~~. 
\label{eq11}
\end{equation}

The resulting regression function are practically coincident with those Eq.~\ref{eq11} 
and $q$ was found so very close to $-0.5$, that we decided to freeze it at this value
leaving only $K(\beta)$ as free parameter.
The final curves and their best fit laws are given in Fig.~\ref{frhoper}.
Note that for both laws the ratio $K(4)/K(3) \approx 4/3$, suggesting that $K(\beta)$, 
at least in this rather narrow range, may be linear, but more calculations are
necessary to unravel this dependence.
We expect, therefore that a slowly modulated $J(t)$ would result in a change of the
the mean recurrence time between spikes.

\section{The origin of low frequency QPOs}
\label{sect:qpo} 

\begin{figure}
\includegraphics[height=8.8cm,angle=-90,scale=1.0]{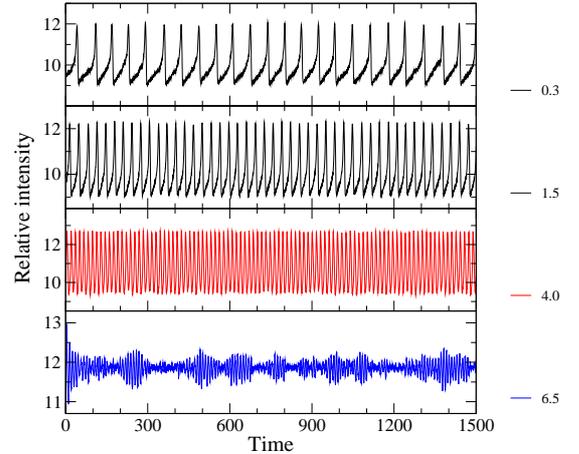}
\caption[]{
Light curves computed applying the MHR mathematical model assuming $J(t)$ of
Eq.~\ref{eq8}, with $C = 3.5$ and different $J_0$ to change the position of the
equilibrium point from the unstable to the stable region.
The values of $J_0$ from top to bottom are 0.3, 1.5 ($\rho$ class), 4.0 (hfo), 
6.5 (QPO).
}
\label{frho2}
\end{figure}

\begin{figure}
\includegraphics[height=8.8cm,angle=-90,scale=1.0]{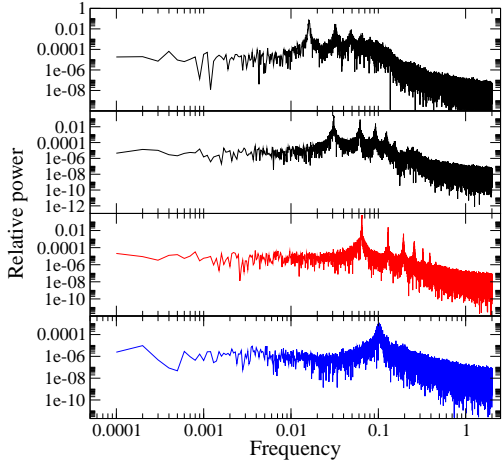} 
\caption[]{Power density spectra of the light curves in Fig.\ref{frho2}. 
The two upper panels correspond to the spiking of the $\rho$ class,
while the third panel (red curve) is that of the `hfo'. 
All these spectra are dominated by the narrow lines and their harmonics
of the periodic signals.
The blue spectrum in the last panel has a unique broad peak typical of
a QPO.
}  
\label{frho3}
\end{figure}

In the previous sections, we showed that the MHR model reproduces the 
sequence of classes $(\phi,\chi) \rightarrow \delta \rightarrow \rho$ for 
increasing values of $J_0$ as illustrated by Fig.~\ref{f2tr}.
When $J_0$ increases, the recurrence time of $\rho$ spikes decreases 
(see Fig.~\ref{frhoper}), and their profiles become more and more symmetrical up 
to approximate a sinusoidal `hfo' pattern, whose amplitude is slightly variable 
because of the fast random changes of $J(t)$.
Some examples of light curves are shown in the three upper panels in 
Fig.~\ref{frho2}.
A further increase of $J_0$ produces a transition to the second stable region 
(see Fig.~\ref{f2tr}) thus we expect a signal rapidly evolving to a constant level, 
however, the occurrence of fast fluctuations of $J_0$ in the unstable region 
determines a random appearance of `hfo' with an amplitude modulation which sets 
on longer time scales (bottom panel in Fig.~\ref{f2tr}).
The PDS (Fig.~\ref{frho3}) of this light curve shows a broad feature, typical of 
QPOs frequently found in binary X-ray sources and, in particular,  
in low mass X-ray binaries hosting a Black Hole or a Black Hole candidate, 
like \grs.
Finally, as discussed previously, when $J_0$ reaches values high enough that the 
equilibrium point remains always in the stable region, the structure of the  
light curve results again that of the $\chi$ class.

\citet{vandenEijnden2016} applied an optimal filtering to the
\grss light curves to obtain the structure of the signal in the frequency band of 
the QPO.
The resulting shape (see Fig.~4 in their paper) is that of a periodic oscillation
whose amplitude is modulated by a non periodic pattern with a time scale higher 
than that of the oscillation by a factor of between about 5 and 10.
It is interesting to note that this filtered signal turns out remarkably similar 
to the one plotted in the bottom panel in Fig.~\ref{frho2}.

On the basis of these results it is possible to formulate an hypothesis on the
origin of QPO in an accretion disc.
They are essentially related to the same mechanism responsible of the spiking 
limit cycle but appears at a transition between the unstable and the stable region 
for high $J_0$.
Random or turbulence fluctuations may be responsible  for the alternance between 
the two regions thus producing the amplitude modulation of the oscillations.

\subsection{Structure of the QPO feature}

\begin{figure}
\includegraphics[height=8.8cm,angle=-90,scale=1.0]{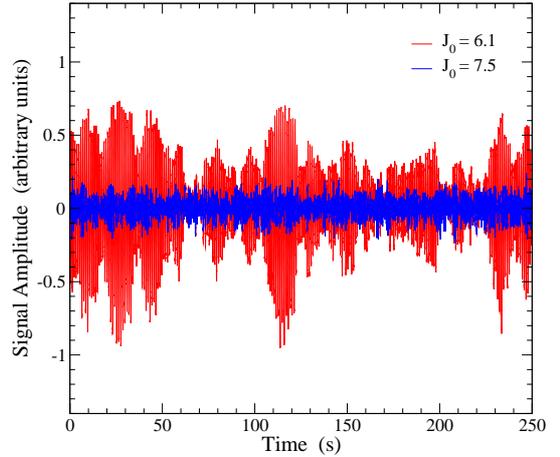}  \\
\includegraphics[height=8.8cm,angle=-90,scale=1.0]{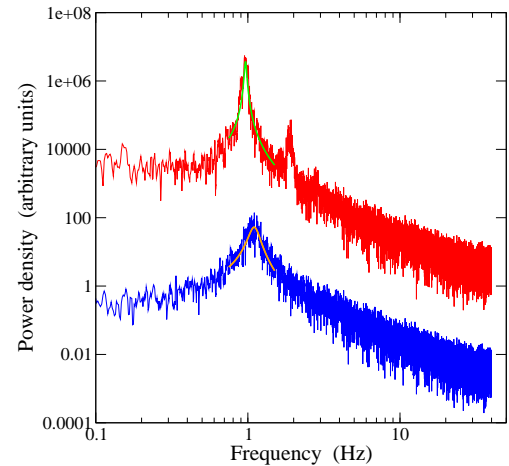} 
\caption[]{
{\it Upper plot}: Light curves computed applying the MHR mathematical model assuming 
$J(t)$ of Eq.~\ref{eq8}, with $C = 3.5$ and two values of $J_0 = 6.1$ (red curve), just
higher than the transition level and 7.5 (blue curve), both located in the stable region.
{\it Lower plot}: power density spectra of these curves, shifted to separate the profiles 
of QPO features.
The green and oranges thick curves are the Lorentzian fits of the QPO peaks.
}
\label{wmc}
\end{figure}

We performed additional numerical calculations to verify how the structure of 
the QPO feature in the PDS depends upon $J_0$.
We considered the two following cases: $i$) a $J_0$ value just above the upper 
boundary of the unstable region (see Sect.~\ref{sct:nullc} and Appendix~\ref{appendix2}), 
namely 6.1, in order to have about 50\% of $J(t)$ values in the unstable region, 
because of the random fluctuations, and $ii$) $J_0 = 7.5$, so that only about 
7\% of them are in this region.
The time scale of these signals was established to have QPO frequencies near 1 Hz.
Two short segments of light curves are reported in the upper plot of Fig.~\ref{wmc} 
from which clearly results the presence of a modulated `hfo' with amplitudes 
depending upon $J_0$ being larger when this quantity is just above the boundary, 
while the other resembles to one of $\chi$ class.
Their PDS are given in the lower plot: large QPO features are in both spectra 
at very close frequencies, namely 0.965 and 1.10 Hz, but the one for $J_0 = 6.1$ 
is more prominent and at least two harmonics are apparent.
Note the high similarities between these results and the spectra reported by
\citet{Stiele2014} for three \rxx observations in the energy band 4.9 - 14.8 keV.
Profiles of both features can be well described by Lorentzian functions from which 
one can estimate their Q factors that result 5.8 and 29.2. 
These values are also dependent on the amplitude of random component, because when 
it is reduced to 0, the light curve, after a rather fast transient, reaches a 
constant.
Of course a further increase of $J_0$ would continue to reduce the value of Q 
up to the complete quenching of the QPO.

\section{Summary and discussion}

It is known from the study of dynamical systems \citep[see, e.g][]{Strogatz1994} 
that the development of convective energy transport and chaotic turbulent motions 
in a fluid can be described by systems of non-linear equations, as e.g. the very 
famous one due to \citet{Lorenz1963}.
In the astrophysical contest \citet{Arnett2011} applied Lorenz equations to 
investigate recurrent fluctuations in turbulent kinetic energy in a stellar oxygen 
burning shell.  

We used a similar approach and derived the MHR model, based on a system of 
two ODEs, whose results have shown that some complex patterns, like those
exhibited by \grs , are well described by non-linear processes and by transitions 
between stable and unstable states driven by a single input function $J(t)$.
Synthetic light curves remarkably very similar to those of $\phi$, $\chi$ and 
$\delta$ classes are obtained in conditions of a stable equilibrium and a constant 
input added to random fluctuations as reasonably expected in a turbulent hot plasma.
The $\delta$ class requires that the mean input would be close to stability
threshold (in our simple case it is close to $J_0 = 0$) and its detailed 
structure depends upon the amplitude and time scales of input fluctuations, 
whose values could exceed the unstable interval boundary.
The MHR equations thus introduce correlations in the output over time scales 
longer than the one of fluctuations and originate a red noise PDS.
The idea of considering a non-linear ODE for describing some properties of accretion
disc oscillations was already adopted by other authors: for instance, \citet{Ortega2014} 
considered the harmonic oscillator equation including quadratic and cubic terms and
a sinusoidal forcing for describing the spectral structure of the normal mode 
oscillations in a thin accretion discs.

An outstanding feature of the MHR model is its ability to produce a spiking state,
again very similar to the $\rho$ class signals, without the introduction of any 
{\it ad hoc} assumption on the time variations of the accretion rate or any other 
physical quantity of the accretion processes.
This capability is also for other simpler ODE systems, like the FitzHugh-Nagumo 
model presented in \citet{Massaro2014}, but the solutions obtained by means of MHR 
model are able to better describe the transitions between some classes and have
therefore a higher heuristic content.
The shape of spikes is well reproduced also in the cases where a dip appears just
after the end of the peak,  characteristic that we used to define the $\rho_d$ subclass.
The condition to produce this pattern is that the fluctuations with respect to a
stable value of the input function move its value across the instability boundary 
and interrupt the completion of the limit cycle making the spike separation 
highly irregular.
Both these transition effects induced by noise can be easily verified by 
eliminating the fluctuations (i.e. fixing $C = 0$) and obtaining purely stable or 
spiking solutions, as shown in Sect.~\ref{sect:rho}.
An important property of the spiking is the dependence of the recurrence time on the
length of the rising segment while the peak width remains practically stable, in a
very good agreement with the observational results \citep{Massaro2014,Massaro2010}.

To make clear how the MHR model can develop the limit cycle of the $\rho$ class
it is useful to consider the single ODE given in Eq.~\ref{eq5:app1}:

\begin{equation}
 \ddot{x} + F(x) \dot{x} + x^3 - J_0 = 0 ~~~~.
\label{eq12}
\end{equation} 

It is easy to verify that, after a multiplication by $\dot{x}$, it can be transformed
into the equivalent form:

\begin{equation}
   \frac{dE}{dt} = - \dot{x}^2 F(x) ~~~~~~,
\label{eq13}
\end{equation}

\noindent
where

\begin{equation}
   E(x,\dot{x}) = \frac{1}{2} \dot{x}^2 + \int (x^3 - J_0) dx  ~~~~~~.
\label{eq14}
\end{equation}

In the case that $F(x)$ is always positive or negative, the quantity 
$E$ will be indefinely decreasing or increasing, respectively; 
these two situations correspond either to the approach to stable equilibrium or 
to an unstable diverging amplitude.
If $F(x)$ has both positive and negative values, one can have an alternance 
between damping and excitation and $E$ and $x$ have an oscillating behaviour, 
that is named `self-excited', because of the absence of any periodic forcing.

In our case $F(x)$ is a second degree polynomial equal to the opposite of the 
trace of the Jacobian (see Appendix  ~\ref{appendix2}), and therefore for 
$\beta^2 > 3$ there are real zeros, then the sign of $F(x)$ can change and 
self-excited oscillations are possible.
For $\beta^2 < 3$ the values of $F(x)$ are all negative and the solution converge
rapidly to equilibrium. 

A  limitation of MHR model, at least in the present version, is that it 
does not take into account the dependence of the spiking profiles on the photon 
energy.
We know from a previous analysis that the full width half maximum of spikes decreases 
with energy following a relationship close to a power law with an exponent $\approx 0.8$ 
\citep{Maselli2018}.
The model light curve for the $\rho$ class shown in Fig.~\ref{frho} has therefore a 
spike FWHM corresponding to an intermediate energy in the considered range,
depending upon the instrumental response.
We did not include any parameters in the MHR model to account for the energy 
dependent properties of the light curves.
A change of the ODE system including more terms and more parameters
will make the stability analysis of solutions much harder than in our case.

A serendipitous and very interesting result of the MHR model presented in 
Sect.~\ref{sect:qpo} is the occurrence of low-frequency QPOs when the system 
reaches the stable region for high $J_0$ values.
For increasing $J_0$ the limit cycle evolves towards the region of `hfo', whose 
frequency is rather stable and changes very little as shown in Fig.~\ref{frhoper}.
QPOs are due to amplitude modulation of these `hfo' on time scales much longer 
than their period.
It is interesting that the model reproduces quite well the amplitude modulation 
resulting in the data applying an optimal frequency filtering as shown by 
\citet{vandenEijnden2016}.
An alternative possibility was proposed by \citet{Sukova2015} who investigated 
the time evolution of shock oscillations by means of numerical hydro-dynamical 
simulations: their solutions for the accretion rate exhibit patterns 
similar to hfo and could be also associated with low-frequency QPOs.
However, a connection of the \citet{Sukova2015} model with the limit cycle
of the $\rho$ class is not established.
According our results, the same processes generating the spiking limit cycles 
around an unstable equilibrium point is also responsible of QPOs, but in this 
case the disc structure remains in a stable condition for most of the time.
This is probably the general condition of the other low mass X-ray binary
accretion discs that generally shows QPOs or red noise.
A more detailed study of non-linear mechasmims for producing QPOs and the
relevance of the noise will be investigated in a further work.

It is useful to distinguish between low and high-frequency QPO.
As written above, our results suggest that the low-frequency QPOs, typically in 
the range from a fraction to a few Hz \citep{Yan2017}, can be related to the 
changes of a noisy input function across the border between unstable and stable
equilibrium region, while high-frequency QPOs can originate by a different 
physical mechanism that likely involves local oscillations and waves in the disc.
Hydrodynamic calculations \citep{Reynolds2009} for a thin accretion disc
have shown that turbulence does excite QPOs whose PDS has a red continuum with a
broad peak at frequencies close to the radial epicyclic frequency at 
$\nu_{re} \approx 10^3 (M_{\odot}/M)$ Hz where $M$ is the central black hole mass.
In the case of \grss the estimated mass of about 12 $M_{\odot}$ \citep{Reid2014}
and the resulting $\nu_{re} \approx 80$ Hz are therefore comparable to those of 
the observed high frequency QPOs \citep{Belloni2013}.

As already noticed above for the $\rho$ class, the present version of MHR 
model is not able to describe the properties of low frequency QPOs depending 
on energy as those reported by various authors 
\citep[e.g.][]{Stiele2014, Zhang2015, Ingram2015}.
Again an extended model with more parameters is necessary, but reasonably
one could expect more complex stability conditions for deriving the various 
light curves of various classes and transitions between them.

An important subject emerging from our modelling is the fundamental role 
of the spectrum of the plasma turbulence in the disc.
Some classes, like $\chi$ and $\delta$, and QPOs are obtained by the MHR 
model only if a noise contribution is considered.
For what concerns the amplitude distribution we made the simplest assumption 
of a random white noise, variable step by step, and with a uniform distribution 
in the interval [$-C/2$, $+C/2$].
A more realistic noise model will have to take into account the probability 
distribution function produced by turbulent motions in the disc, as for 
instance a log-normal law that, considering the energy dissipation in Alfvenic 
turbulence, well describes the statistical properties of this phenomenon 
\citep[see][]{Zhdankin2016}.
The relevance of turbulence in disc dynamics has been recently investigated by 
\citet{Ortega2020}, who pointed out that stochastic oscillations can behave as a 
driving agent for producing the twin peak structure of high frequency QPOs.

Light curves obtained by means of the MHR model are not limited to those 
described in the present paper which were all computed using a constant $J_0$.
It is possible to show that structures as those of several other classes 
(for instance $\alpha$, $\gamma$, $\lambda$, $\kappa$, $\omega$, $\xi$, 
$\theta$) result when this assumption is  relaxed and a step function 
or sawtooth modulations of $J(t)$ on the proper time scales are introduced.
The extension of MHR model to other variability classes will be discussed 
in detail in \citetalias{Massaro2020}.

The physical meaning of the input function $J(t)$ is likely related to some 
parameter affecting the disc state and consequently its brightness and 
stability.
In the paper concerning the FitzHugh-Nagumo model \citet{Massaro2014}
proposed that it is related to the mass accretion rate in the disc.
This hypothesis remains useful also in the context of the MHR model but it 
requires a further analysis to establish a reliable functional form.
In particular, one could use the large collection of observation to search
for correlations between the properties of the various variability classes 
and the mean luminosity and its changes or some other spectral parameter.
This draining work is beyond the goals of the present paper that is focused 
on the development of a tool for simulating the stability conditions that 
produce the rich collection of variability classes.

A more detailed comparison between our MHR model and these disc instabilities 
will be given in \citetalias{Massaro2020}.
 
\section*{Acknowledgments}
The authors are grateful to Enrico Costa, Marco Salvati and Andrea Tramacere 
for their fruitful comments.
We are also grateful to the referee M. Ortega-Rodriguez for his constructive
comments and suggestions.
MF, TM and FC acknowledge financial contribution from the agreement ASI-INAF 
n.2017-14-H.0



\bibliographystyle{mnras}
\bibliography{grs1915} 

\appendix

\section{The differential equation for \emph{x}}
\label{appendix1}

The ODE system of Eq.~\ref{eq3}, or that of Eq.~\ref{eq5}, can be reduced to a 
single ODE for the variable $x$ by eliminating $y$ and $\dot{y}$; deriving the 
first ODE in Eq.~\ref{eq3}, we have:
\begin{equation}
 \ddot{x} = (- 3 \rho x^2 + 2 \beta_1 x) \dot{x} + \dot{y} + \dot{J}(t)  \nonumber  
\label{eq1:app1}
\end{equation} 
and using the equations for $\dot{y}$, it follows
\begin{equation}
 \ddot{x} = (- 3 \rho x^2 + 2 \beta_1 x) \dot{x} - \beta_2 x^2 - y + \dot{J}(t)  ~~~~.\nonumber
 \label{eq2:app1}
\end{equation} 

Then one can use the ODE for $\dot{x}$ to eliminate $y$, and the final result is
\begin{equation}
 \ddot{x} + F(x) \dot{x} - (\beta_1 - \beta_2) x^2 + \rho x^3 - [J(t) + \dot{J}(t)] = 0 
  \label{eq3:app1}
\end{equation} 
where
\begin{equation}
 F(x) = 3 \rho x^2 - 2 \beta_1 x + 1    ~~~~. 
 \label{eq4:app1}
\end{equation} 

In the simple case with $\beta_1 = \beta_2$, $J(t) = J_0$, and
$\rho = 1$ (with no loss of generality), it becomes:
\begin{equation}
 \ddot{x} + F(x) \dot{x} + x^3 - J_0 = 0 ~~~~.
\label{eq5:app1}
\end{equation} 

Another simple derivation of this equation is obtained by introducing the new variable:

\begin{equation}
 w(t) = x(t) + y(t)  ~~~;~~~  \dot{w} = \dot{x} + \dot{y}  ~~~ \nonumber
 \label{eq6:app1}
\end{equation}

\noindent
and adding the two equations for $\dot{x}$ and $\dot{y}$ we have:

\begin{eqnarray}
\frac{dx}{dt} &=& - x^3 + \beta~ x^2 - x + w + J_0 \nonumber \\
\frac{dw}{dt} &=& - x^3 + J_0  
\label{eq7:app1}
\end{eqnarray} 

This ODE system is equivalent to the MHR of Eq.~\ref{eq3}, but the second variable
is different.
Deriving the equation for $\dot{x}$ and using that for $\dot{w}$ we obtain
again the previous Eq.~\ref{eq5:app1}.

\section{ Nullclines, equilibrium points and instability interval}
\label{appendix2}

Let us consider the model of Eq. (3), with $\rho = 1.0$ without loss
of generality

\begin{eqnarray}
\frac{dx}{dt} &=& - x^3 + \beta_1 x^2 + y + J(t) \nonumber \\
\frac{dy}{dt} &=& - \beta_2 x^2 - y  
\label{eq3:app2}
\end{eqnarray}

It is possible to demonstrate that if $J(t)$ is a bounded function, 
$|J(t)| \le M, \forall t$, also the solutions for $x(t)$ and $y(t)$ are bounded.

{\it Lemma}: If $J(t)$ is bounded then exists a number $b > 0$, such that for 
any initial condition $(x(0)$, $y(0))$ it exists a value $\bar{t} > 0$ such that 
$\Vert x(t)$, $y(t) \Vert \le b$, $\forall t > \bar{t}$.

{\it Proof}: Let $ V(x,y,t) = (1/2)(x^2 + y^2)$ then 

$\dot{V} = -x^4 + \beta_1 x^3 + xy + x J(t) - \beta_2 x^2 y - y^2 \le G(x,y)$

\noindent
where 

$G(x,y) = -x^4 + \beta_1 x^3 + xy - \beta_2 x^2 y - y^2 + |x| M$,

\noindent
and using polar coordinates it is easy to verify that 
{\setlength{\mathindent}{0cm}
\begin{flalign}
\lim_{x^2 + y^2 \rightarrow +\infty} G(x,y) = -\infty ~~.
\nonumber
\end{flalign}}

To study the stability conditions of the ODE system in Eq. (4) 
with $J(t) = J_0$ we consider the system

\begin{eqnarray}
\frac{dx}{dt} &=& - x^3 + \beta~ x^2 + y + J_0 \nonumber \\
\frac{dy}{dt} &=& - \beta x^2 - y  
\label{eqA2-1}
\end{eqnarray}
 
\noindent
Thus its {\it nullclines} are given by the equations: 

\begin{eqnarray}
 y &=&  x^3 - \beta~ x^2 - J_0 \nonumber \\
 y &=& - \beta~ x^2  
\label{eqA2-2}
\end{eqnarray}

\noindent
which admits the unique real equilibrium point 
$x_* = J_0^{1/3}$, $y_* = -\beta~J_0^{2/3}$.
To investigate local the stability of this solution we perform the linear 
analysis at $x_*$ by means of the sign of the determinant and trace of 
the Jacobian:

\begin{displaymath}
\left( \begin{array}{cc}
-3 x_*^2 +2 \beta x_* & 1  \\
-2 \beta x_* & -1  
\end{array} \right)
\label{eqA2-3}
\end{displaymath}

\noindent
whose determinant $\Delta$ is
\begin{equation}
 \Delta =  3 x_*^2 - 2 \beta x_* + 2 \beta x_* = 3 x_*^2 \ge 0 
\label{eqA2-4}
\end{equation}
is non-negative
while the sign of the trace is given by the roots of:
\begin{equation}
 - 3 x_*^2 + 2 \beta~ x_* - 1 = 0 
 \label{eqA2-5}
\end{equation}
\noindent
which are $x_{*1,2} = (\beta \pm \sqrt{\beta^2 - 3}~)/3$.
The local unstable equilibrium condition $Tr > 0$ corresponds to values of $x_*$
within the interval
$I \equiv [(\beta - \sqrt{\beta^2 - 3}~)/3$, $(\beta + \sqrt{\beta^2 - 3}~)/3]$.

For a general analisys of the stability one can introduce the Lyapunov function
\begin{equation}
 V(u,v) = (u^2/4)  (u^2 + 4 x_* u + 6 x_*^2) + (v^2/2)  \nonumber
 \label{eqA2-6}
\end{equation}
\noindent
where $u = x - x_*$,  $v =  (x - x_*) + (y - y_*)$. 
It is easy to verify that: $i$) for $\beta^2 \le 3.0$ 
and $x_* \notin I$, one has $\dot{V} \le 0$ and 
therefore the equilibrium is globally stable;
$ii$) for $\beta^2 > 3.0$, and $x_* \le \xi_1$ or $x_* \ge \xi_2$,
with $\xi_{1,2} = (\beta \pm 2 \sqrt{\beta^2 - 3}~)/3$,
one has again $\dot{V} \le 0$ and the equilibrium is globally stable.
For $x_* \in I$, considering that the solution is bounded, one has at
least a periodic orbit.

In the case of $x_1 \lt x_* \lt x_2$ we can prove that if $\beta^2 \gt 3$,
and if $x_*^3 = (1/27) \beta ( 2 \beta - 3)$ then the periodic orbit is
orbitally stable and therefore it is unique 
\citep[see][]{Coppel1965, HaleKojak1991}.

\bsp	 
\label{lastpage}
\end{document}